\documentclass[journal]{IEEEtran}
\pdfoutput=1
\usepackage[pdftex]{graphicx}
\usepackage{amsmath}
\usepackage{url}

\begin{document}
\title{High resolution image reconstruction with constrained, total-variation minimization}
%
%

\author{Emil~Y.~Sidky,
        Rick~Chartrand,~\IEEEmembership{Member,~IEEE}
        Yuval~Duchin,
        Christer~Ullberg,~\IEEEmembership{Member,~IEEE}
        and~Xiaochuan~Pan,~\IEEEmembership{Fellow,~IEEE}
\thanks{Manuscript received November 21, 2010.
This work was supported in part by NIH R01 grants CA120540 and EB000225.
The contents of this article are
solely the responsibility of the authors and do not necessarily
represent the official views of the National Institutes of Health.
This work was also supported in part by the U.S. Department of Energy
through the LANL/LDRD Program.}
\thanks{E. Y. Sidky, Y. Duchin, and X. Pan are with the University of Chicago
Dept. of Radiology, Chicago, IL 60637 USA.}%
\thanks{R. Chartrand is with Los Alamos National Laboratory, Los Alamos, NM 87545 USA.}%
\thanks{C. Ullberg is with XCounter AB, Danderyd, SE-182 33 Sweden.}%
}

\maketitle
\pagestyle{empty}
\thispagestyle{empty}

\begin{abstract}
This work is concerned with applying iterative image reconstruction, based on constrained total-variation
minimization, to low-intensity X-ray CT systems that have a high sampling rate.  Such systems
pose a challenge for iterative image reconstruction, because a very fine image grid is needed to
realize the resolution inherent in such scanners. These image arrays lead to under-determined
imaging models whose inversion is unstable and can result in undesirable artifacts and noise patterns.
There are many possibilities to stabilize the imaging model, and this work proposes a method
which may have an advantage in terms of algorithm efficiency. The proposed method introduces
additional constraints in the optimization problem; these constraints set to zero high spatial
frequency components which are beyond the sensing capability of the detector. The method
is demonstrated with an actual CT data set and compared with another method based on projection
up-sampling.
\end{abstract}


\section{Introduction}

\IEEEPARstart{T}{his} proceedings focuses on a fundamental issue of resolution in iterative image reconstruction (IIR).
IIR is being considered for application in computed tomography (CT), see for example \cite{ASIR:10},
because it is possible to account for noise in the data model and
accordingly allow for high quality images with a reduced exposure \cite{McCollough:09}.
IIR methods being considered for CT, all
involve implicit solution, see section 15.3 of \cite{Barrett:FIS},
for the image as opposed to explicit solution, derived from exact or approximate inverses
to the continuous cone-beam transform.  For explicit reconstruction algorithms, which are generally some variant of
filtered back-projection (FBP), the reconstructed volume can be obtained one point at a time. Images can be
shown on grids of any size and with arbitrarily small
grid spacing, for example blowing up a region-of-interest (ROI).
Note, that does not mean the resolution is arbitrarily high, because the system resolution is still limited
by the discrete data sampling. Reconstruction by implicit solution allows more flexible and realistic data
models for the tomographic system, but at a price. As pointed out often, IIR algorithms
are generally more computationally intensive. Another important issue is that a complete expansion set
for the imaged volume is necessary in order to obtain the reconstruction, and the complete image must be
solved for all at once; the image cannot be gotten voxel-by-voxel.

More concretely, let us consider a linear imaging model using a voxel expansion of the volume
and ideal conditions of perfect data consistency:
\begin{equation}
\label{discreteModel}
\mathbf{g} = \mathcal{X} \mathbf{f},
\end{equation}
where $\mathbf{g}$ represents the projection data as a 1D vector; $\mathbf{f}$ is a vector of voxel values;
$\mathcal{X}$ is the system matrix, which yields line integrals through the volume and for
the present work is based on the ray intersection length as specified in Siddon's method.
This model, which forms the basis of many IIR algorithms, is difficult to solve explicitly
for realistic size models of a CT system. The matrix $\mathcal{X}$ is ill-conditioned and
can be extremely large, up to a size of $10^9 \times 10^9$. Equation (\ref{discreteModel}) is usually
solved implicitly by one of many algorithms, conjugate gradients,  algebraic reconstruction techniques (ART),
etc. Two practical issues arise in the implicit solution of Eq. (\ref{discreteModel}): (I) the voxel
representation of the image must include the whole region where the measured transmission rays intersect
with the support of the subject \cite{ziegler2008iterative}, and
(II) to attain the intrinsic resolution of the data the voxels must
be small compared with the detector bin width. The combination of these two factors causes a computational
burden, because straight-forward voxelization of the reconstruction volume will lead to a very large array.
But also, mathematically, the second point that voxels should be small compared to the detector bins leads
to an under-determined linear system for Eq. (\ref{discreteModel}), because there will be more unknown
voxel values than known ray-transmission measurements. In this work, we examine a couple of possible
solutions for addressing the second issue -- dealing with the large null-space of Eq. (\ref{discreteModel})
in a controlled way.

In Sec. \ref{sec:theory} we present the image reconstruction algorithms and in Sec. \ref{sec:results}
we demonstrate them on an actual low-intensity CT scan.

\section{Constrained, total-variation minimization theory and algorithms}
\label{sec:theory}

In recent years, we have been investigating the solution of Eq. (\ref{discreteModel}), by solving
the constrained TV-minimization problem:
\begin{equation}
\label{TVmin}
\mathbf{f}^* = \text{argmin} \| \mathbf{f} \|_\text{TV} \text{  such that  }
|\mathcal{X} \mathbf{f} - \mathbf{g}|^2 \le \epsilon^2
\; \; \; \mathbf{f} \ge 0,
\end{equation}
where $\| \mathbf{f} \|_\text{TV}$
is the sum over the gradient magnitude image; and $\epsilon$ is a data-error tolerance parameter
necessary because the projection data is likely not consistent with the image model. To solve
this optimization problem, we have been developing a heuristic algorithm
\cite{SidkyTV:06,sidky2008image,sidky2009enhanced,PanIP:09,SidkyPC:10,sidky2010lowintensity}
that alternates between
projection-onto-convex-sets (POCS), to enforce the constraints, and steepest descent (SD) to
reduce image TV. The key point of this algorithm is that the SD step-size is controlled adaptively, so
as to not undo progress toward the feasible image set with POCS. This alternating algorithm is
called adaptive SD-POCS (ASD-POCS). For investigations where solving Eq. (\ref{TVmin}) accurately
is important, optimality conditions should be checked \cite{sidky2008image,sidky2010lowintensity}. On the other hand,
for practical applications it may not be necessary to have an accurate solver \cite{sidky2009enhanced}, and
in such cases ASD-POCS yields a clinically useful image within 10-20 iterations.

Image reconstruction algorithms for CT based on Eq. (\ref{TVmin}) have been shown to be
effective for sparse-view projection data
\cite{SidkyTV:06,song2007sparseness,chen2008prior,sidky2008image,sidky2009enhanced,jia2010gpu,SidkyPC:10,Bergner:10,Choi:10,Bian:10},
which has obvious
implications for patient dose.  More recently, we have been interested in how to apply ASD-POCS
to what has traditionally accepted as fully angularly-sampled data.  For such data, structures
on the scale of a detector bin-width are expected to be resolved.  For IIR,
this requirement poses the above-mentioned problem that the number of voxels may be much larger
than the number of data, and direct application of the IIR algorithm may lead to strange noise
patterns which can interfere with imaging tasks \cite{sidky2010lowintensity}.
There are likely many
ways to resolve this problem; for example, introducing a non-zero cross-section to the ray
model in $\mathcal{X}$ may yield nicer noise patterns while maintaining image resolution.
At the same time, more realistic system model ling usually comes at the expense of computational
speed. As a result, we have been seeking other alternatives.

In Ref. \cite{sidky2010lowintensity},
we propose to stabilize the problem in Eq. (\ref{TVmin}) by making the
following observation: CT resolution is non-uniform, and generally, the angular-sampling is
worse than detector resolution.  Assuming that the individual projection sampling satisfies Nyquist
frequency, each projection can be up-sampled to increase the number of measurement rays at each
projection.  We will still have a data set which is under-sampled in the angular direction with
respect to the very high resolution imaging grid. But we already know that TV-minimization
approaches appear to be effective with this type of under-sampling. We refer to this approach
as up-sampling ASD-POCS or, in this text, as simply up-sampling.

In this proceedings, we propose another approach which may have even a greater advantage in
terms of algorithm efficiency. The method comes from realizing that while we need many
voxels per bin-width for flexibility of the image representation, we cannot hope to actually
attain true super-resolution where structures smaller than the detector bin are visible.
To capture this idea mathematically, we impose constraints on the Fourier transform of the image.
Specifically, high frequency components of the image are set to zero:
\begin{equation}
\label{freqConstraint}
\mathcal{F} \mathbf{f} (| \mathbf{\nu} |>\nu_\text{max})= 0,
\end{equation}
where $\mathcal{F}$ is the discrete Fourier transform (DFT); $\mathbf{\nu}$ represents spatial
frequency; and the frequency magnitude $\nu_\text{max}$ is determined by the bin spacing.
If the detector bins have width $w$, we set $\nu_\text{max}=1/(2 \pi w)$. Although this choice
makes sense physically, $\nu_\text{max}$ can be taken as a algorithm parameter, but
varying $\nu_\text{max}$ is beyond the scope of this study. Adding this constraint
to the optimization problem in Eq. (\ref{TVmin}) leads to frequency-constrained ASD-POCS.
This approach involves processing fewer data in exchange for computing DFTs, which can
be done efficiently with the fast Fourier transform. The pseudo-code for frequency-constrained
ASD-POCS is the same as that reported in Ref. (\cite{sidky2010lowintensity})
with the additional
steps of enforcing Eq. (\ref{freqConstraint}) at each line where the image positivity is enforced.
For the present results, the frequency constraint was imposed before positivity.

\section{Results}
\label{sec:results}

\begin{figure}[!t]
\centering
\includegraphics[width=3.45in]{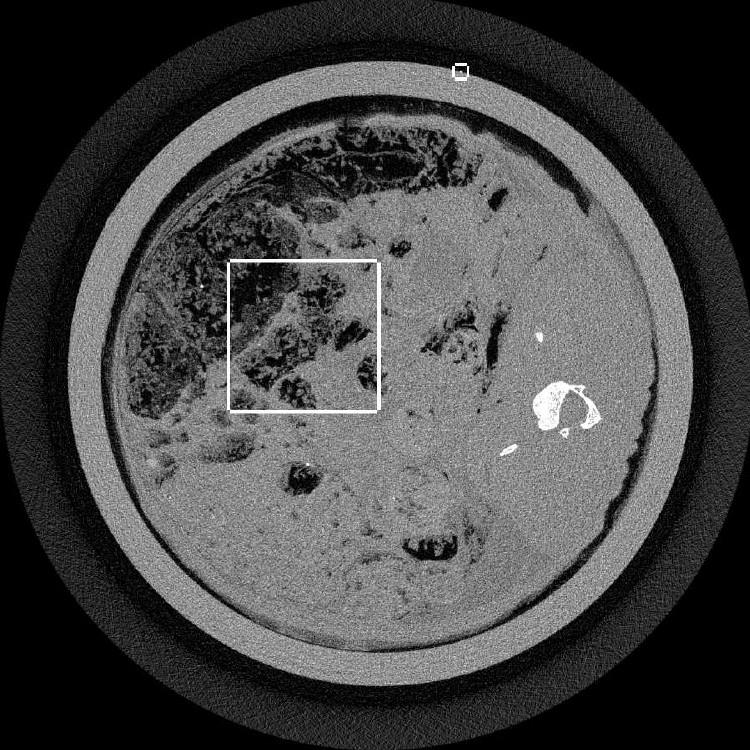}
\caption{FBP slice image of a rabbit scanned by a low-dose XCounter CT system.
The raw FBP image was smoothed by a Gaussian filter, reducing the image TV by a factor of 8.
The indicated rectangles show the ROIs which are used for the algorithm comparisons.
In particular, the small one surrounds the wire object used to obtain a sense of resolution.
The display window is [0,0.04] mm$^{-1}$. }
\label{fig:FBP}
\end{figure}

The up-sampling and frequency constrained ASD-POCS algorithms are demonstrated 
with an XCounter CT scan of a rabbit with a thin wire taped to
the outside of the sample holder. The data are low-intensity and contain 1878 projections with a 2266x64
bin detector at a resolution of 0.1 mm. The thin wire provides a good resolution
test for the image reconstruction algorithm.
For the present purpose,
 we take the middle row on the detector from this data set and focus on 2D fan-beam CT reconstruction
with 1878 projections on a 2266-bin linear detector array.

\begin{figure}[!t]
\centering
Gaussian-filtered FBP
\vskip 0.1cm
\includegraphics[width=3.45in]{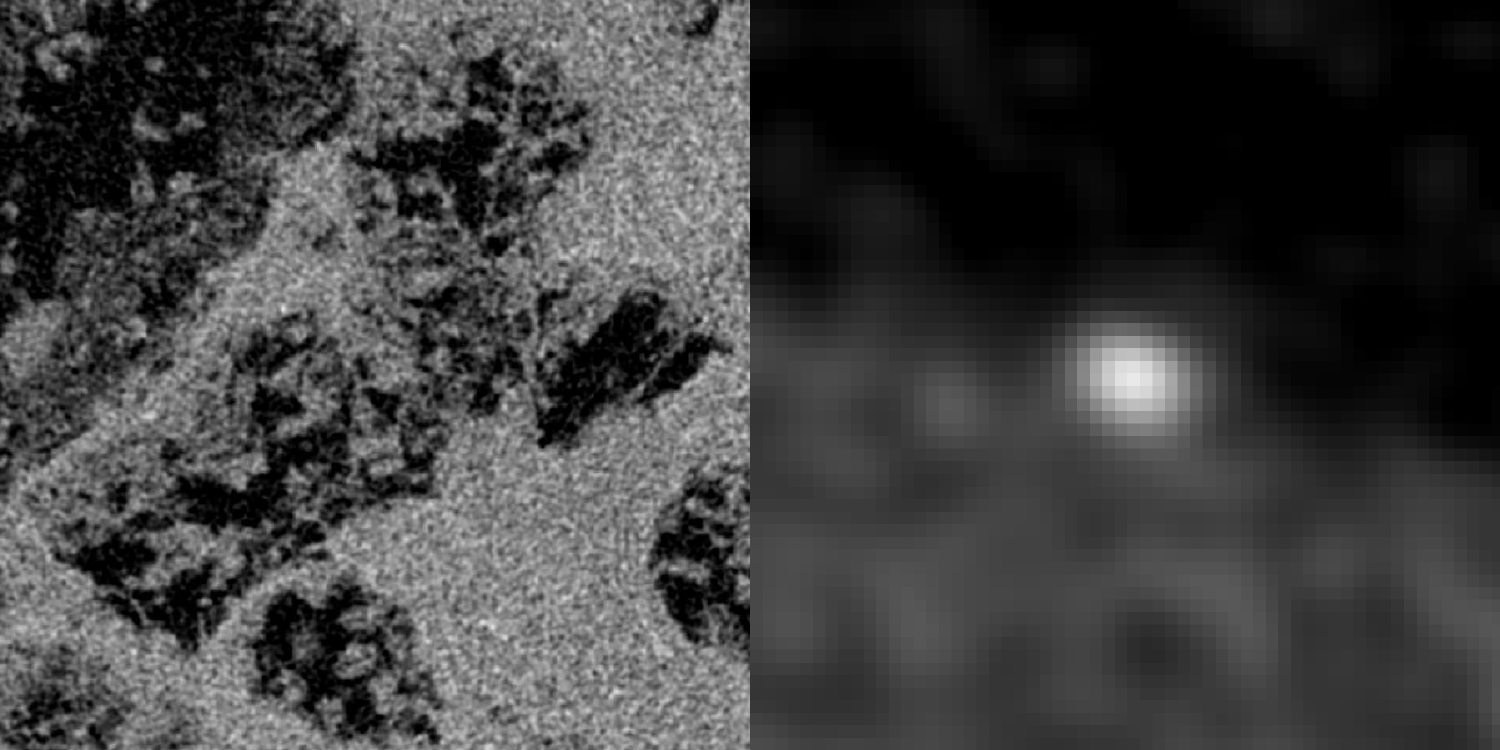}
\vskip 0.5cm
ASD-POCS (0.1 mm pixels)
\vskip 0.1cm
\includegraphics[width=3.45in]{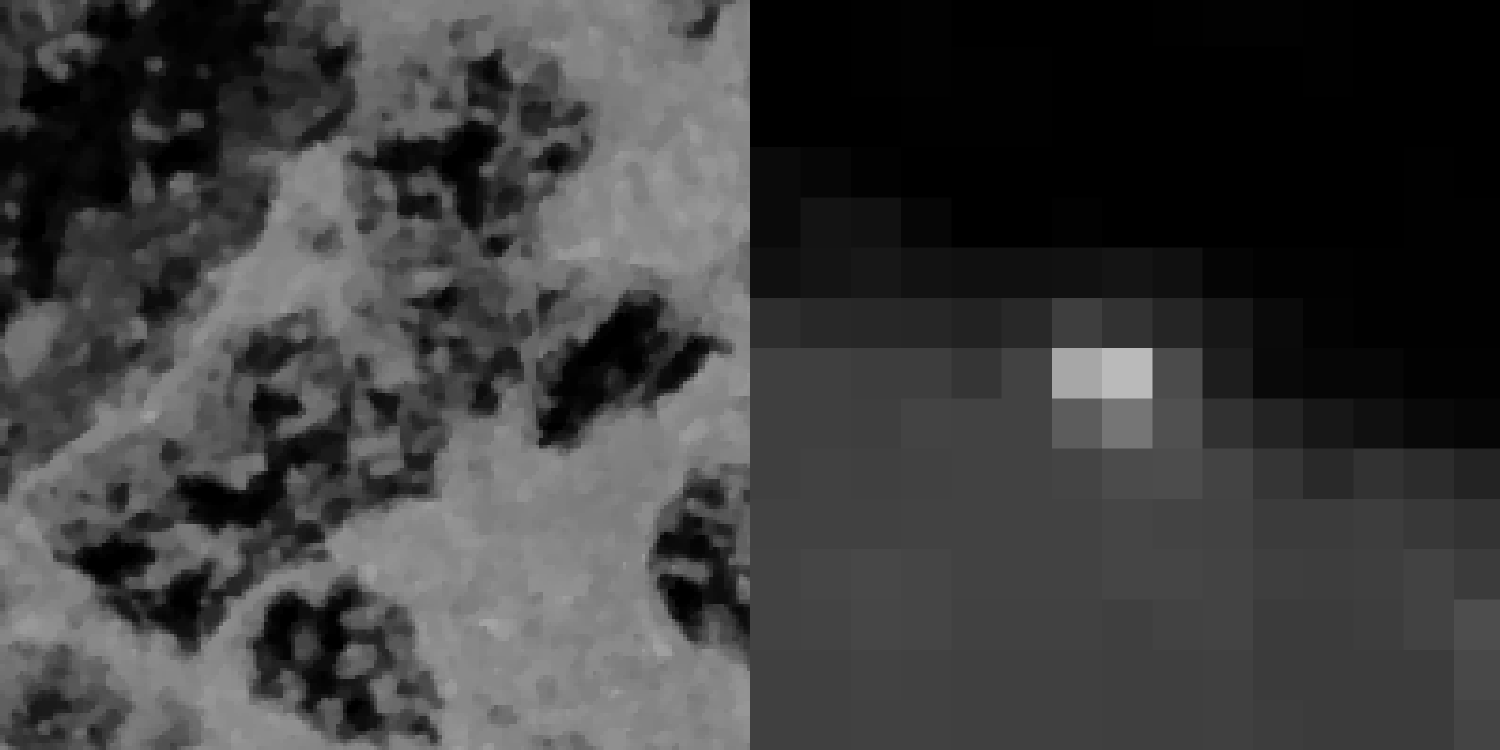}
\vskip 0.5cm
ASD-POCS (0.025 mm pixels)
\vskip 0.1cm
\includegraphics[width=3.45in]{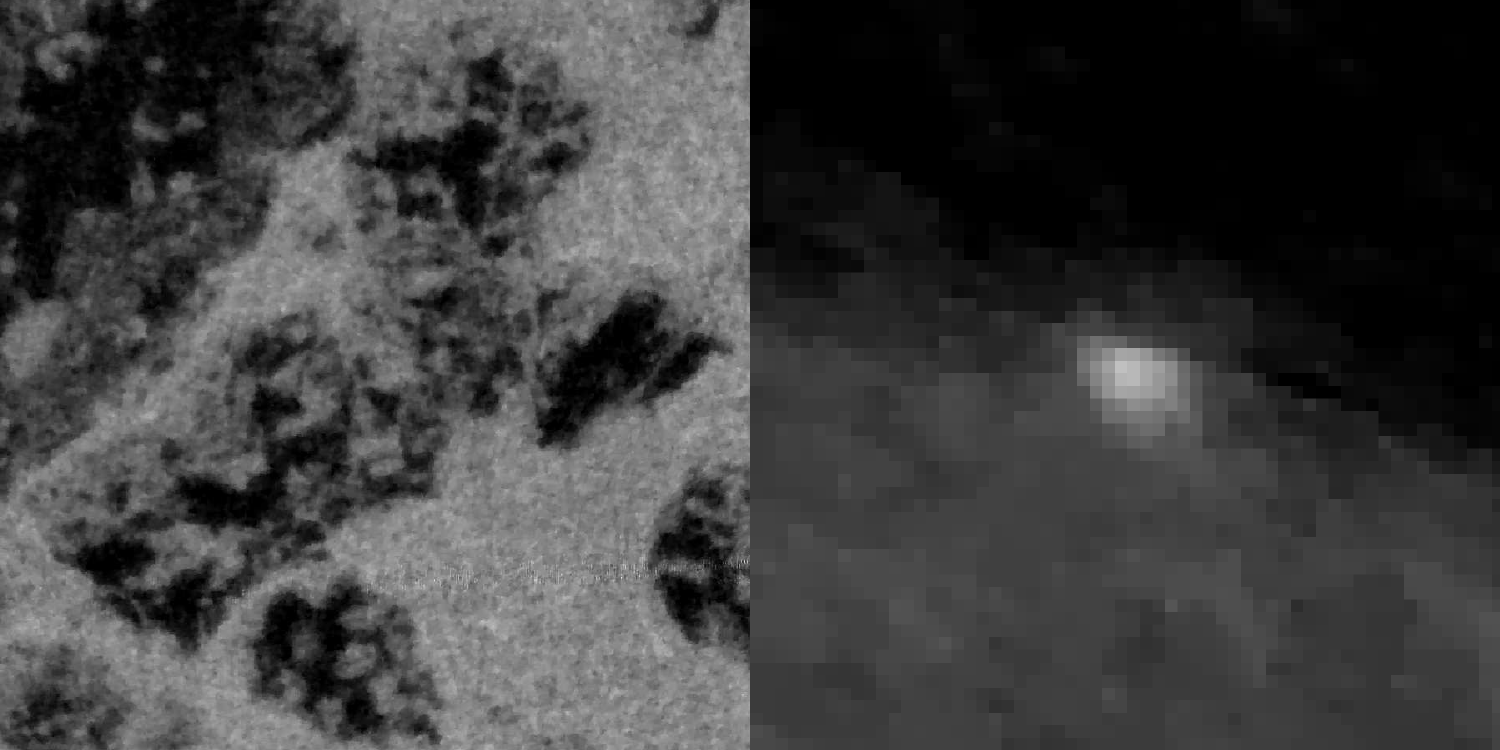}
\caption{(Top) FBP ROI's for the image presented in Fig. \ref{fig:FBP}.
(Middle) ASD-POCS image reconstructed from original data for a grid with 0.1 mm pixels.
(Bottom) ASD-POCS image reconstructed from original data for a grid with 0.025 mm pixels.
The image TV for each case is set to an eighth of that of the raw FBP image.
The display window is [0,0.04] mm$^{-1}$ on the left ROI and [0,0.09] mm$^{-1}$ on the right ROI. }
\label{fig:problems}
\end{figure}

An FBP image of this data set is shown in Fig. \ref{fig:FBP}, where some regularization is performed
by Gaussian smoothing with a window 2 pixels wide. The rectangles indicate the regions where comparisons
of the different algorithms are shown.  Comparisons for each algorithm will be made at a level of image
regularization where each image's TV-norm is an eighth of that of the unregularized FBP image. The images
are shown just to give a sense about the behavior of the algorithms; more rigorous evaluation with
different levels of regularization will
be performed in future work.

We illustrate the problems with employing a matched-resolution
grid, 1024$\times$1024, and a very high resolution grid, 4096$\times$4096,
with the basic ASD-POCS algorithm.
The pixel width is 0.1 mm for the former grid and 0.025 mm for the latter.
Note that the rabbit support projects
onto the middle 1000 bins of the full projection.
The FBP image is shown on a 4096$\times$4096 grid, but because FBP is based on an explicit inverse
its pixel values are not affected by the grid size.
In Fig. \ref{fig:problems} the matched grid clearly leads to a loss of resolution relative to
FBP, because the image is forced to be uniform over the 0.1mm $\times$ 0.1mm squares.
Simply going to a larger array does improve the ASD-POCS image, but the noise pattern
shows additional false structure when compared with the FBP image. These structures
arise from the fact that the imaging problem is under-sampled by roughly
a factor of ten and this under-sampling occurs in both view-angle and detector-bin
directions. The random, sparse specks that
appear could be confused with micro-calcifications in the context of breast imaging.

\begin{figure}[!t]
\centering
ASD-POCS (projection up-sampling)
\vskip 0.1cm
\includegraphics[width=3.45in]{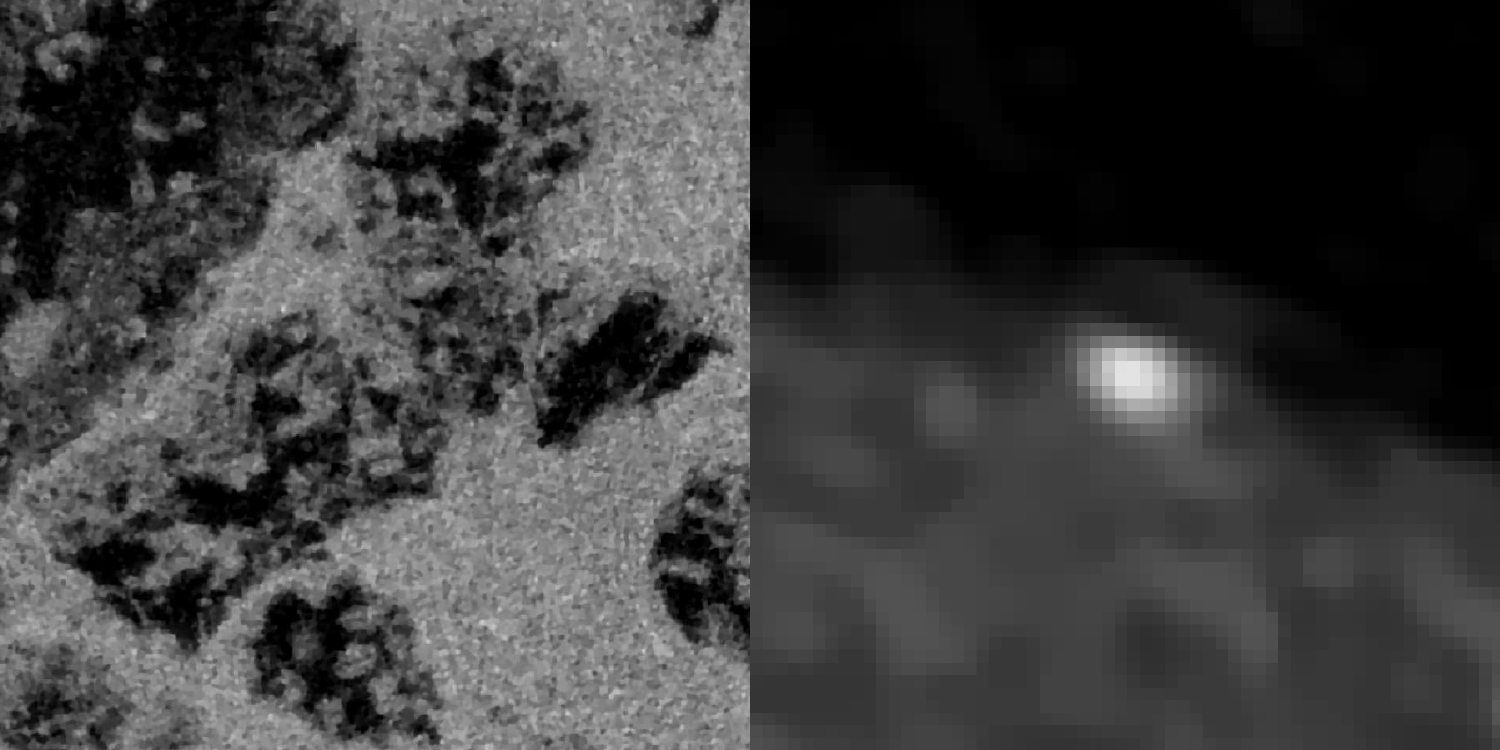}
\vskip 0.5cm
ASD-POCS (spatial frequency constraints)
\vskip 0.1cm
\includegraphics[width=3.45in]{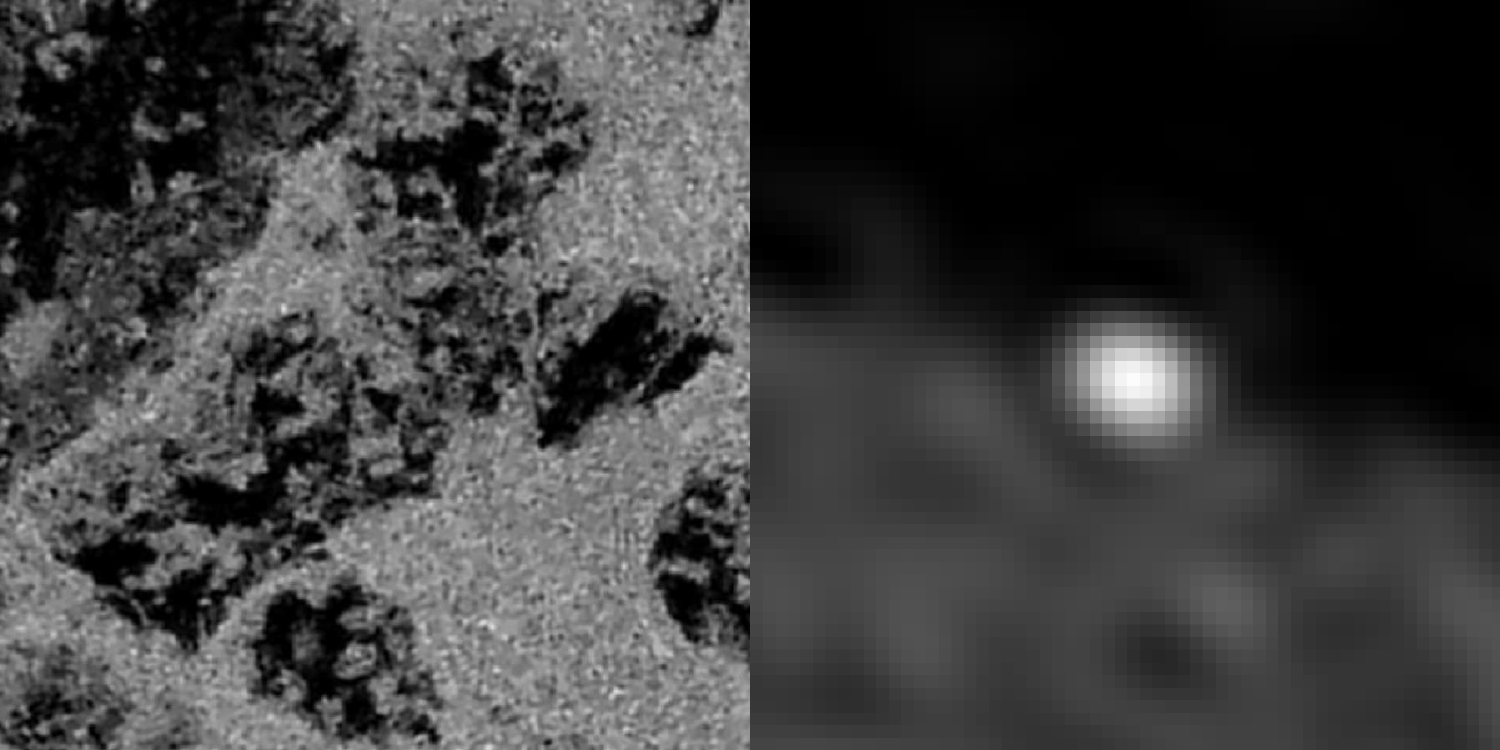}
\caption{(Top) ASD-POCS image obtained from the up-sampled data set.
(Bottom) ASD-POCS image reconstructed from original data set with additional constraints
on the spatial frequency.
For both images the pixel width is 0.025 mm.
The image TV for each case is set to an eighth of that of the raw FBP image.
The display window is the same as that of Fig. \ref{fig:problems} }
\label{fig:newASDPOCS}
\end{figure}

Finally, we show images for ASD-POCS using up-sampling and frequency constraints,
and the resulting region-of-interest images are shown in Fig. \ref{fig:newASDPOCS}.
For the up-sampling method the data are up-sampled at each projection so that the
data set size nominally becomes 1878 views by 9064 virtual bins prior to reconstruction
by ASD-POCS. The frequency constraint ASD-POCS takes the original 1878$\times$2266
data set as input.  As can be seen in the figure, both approaches remove the disturbing
noise pattern seen at the bottom of Fig. \ref{fig:problems}.  Each of these images
show some potential advantage over FBP in that the wire appears to be better focused in the up-sampling
approach,
and the image noise level is reduced approximately 20\% for both up-sampling and frequency
constrained ASD-POCS.  It is possible that the frequency constraint method could lead to better resolution
by allowing $\nu_\text{max}$ to be increased. Such a study will be investigated in future work.

\section{Conclusion}

We have developed modifications to ASD-POCS that allow for high-resolution recovery of
structures occurring on the scale of the detector bin width.  In particular, in this work
we show that including constraints on the spatial frequencies of the reconstructed image
can improve ASD-POCS images by eliminating false structures, which arise from the large
null-space inherent in the imaging model when very high-resolution grids are used
to represent the image.  Further research, will compare the up-sampling
and frequency constraint approaches with many data sets and different levels of
image regularity, and we will also look into implementing more
realistic ray-models that account for source spot-size and the extent of the detector bins.
This realistic modeling in conjunction with frequency extrapolation \cite{chartrand2010}
may allow even further improvement of CT image quality.


\bibliographystyle{IEEEtran}
\bibliography{lowdose}
%





\end{document}